\documentclass[12pt,letterpaper]{article}  

\usepackage[includeheadfoot,
            marginratio={1:1,2:3}, 
            width=430pt, 
            height=688pt,]{geometry}

\usepackage{verbatim}
\usepackage{amsmath}
\usepackage{amsfonts}
\usepackage{amssymb}
\usepackage{mathtools}
\usepackage{physics}
\usepackage{graphicx}
\usepackage{tikz} 
\usetikzlibrary{decorations.pathreplacing,calc,calligraphy}
\usepackage{booktabs}
\usepackage{adjustbox}
\usepackage[utf8]{inputenc}
\usepackage{multirow}
\usepackage[splitrule]{footmisc}

\usepackage{empheq}
\usepackage{paralist}
\usepackage{cite}
\usepackage[hidelinks]{hyperref}
\usepackage{xcolor}
\usepackage{tensor}

\newcommand{\nc}{\newcommand}
\nc{\lb}{\llbracket}
\nc{\rb}{\rrbracket}
\nc{\gl}{\llbracket}
\nc{\gr}{\rrbracket}
\nc{\del}{\partial}
\nc{\tri}{\hspace{-3.5pt}\vartriangle\hspace{-3.5pt}}
\nc{\blacktri}{\blacktriangle}

\nc{\eq}[1]{\begin{equation}
                     \begin{split} #1 \end{split}
                     \end{equation}}
\nc{\ul}{\underline}
\nc{\ov}{\overline}

\nc{\fa}{\hat}
\nc{\fb}{\MakeUppercase}
\nc{\fc}{\tilde}
\nc{\Lie}{{\cal L}} 
\nc{\lambdabar}{{\mkern0.75mu\mathchar '26\mkern -9.75mu\lambda}}

\allowdisplaybreaks[2]
\numberwithin{equation}{section}

\DeclareUnicodeCharacter{2212}{-}
\begin{document}

\vspace*{-1.5cm}
\begin{flushright}
  {\small
  MPP-2022-58\\
  }
\end{flushright}

\vspace{0.5cm}
\begin{center}
{\LARGE
  Mass Hierarchies and Quantum Gravity \\[0.4cm]
  Constraints in DKMM-refined KKLT  
} 
\vspace{0.2cm}

\end{center}

\vspace{0.15cm}
\begin{center}
Ralph Blumenhagen$^{1}$,
Aleksandar Gligovic$^{1,2}$,
Seyf Kaddachi$^{1,2}$, \\[0.2cm]
\end{center}

\vspace{0.0cm}
\begin{center} 
\emph{
$^{1}$ 
Max-Planck-Institut f\"ur Physik (Werner-Heisenberg-Institut), \\ 
F\"ohringer Ring 6,  80805 M\"unchen, Germany } 
\\[0.1cm] 
\vspace{0.15cm} 
\emph{$^{2}$ Ludwig-Maximilians-Universit{\"a}t M\"unchen, Fakult{\"a}t f{\"u}r Physik,\\ 
Theresienstr.~37, 80333 M\"unchen, Germany}\\[0.1cm]
\vspace{0.3cm}
\emph{E-mail:} {\small blumenha@mpp.mpg.de,
  a.gligovic@physik.uni-muenchen.de, }\\[-0.15cm]
\hspace{-4.5cm}{\small seyfkaddachi@gmail.com}\\[0.1cm]
\end{center}

\vspace{0.3cm}


\begin{abstract}
  We carefully revisit the mass hierarchies for the KKLT scenario
  with an uplift term from an anti D3-brane in a strongly
  warped throat. First, we derive the bound resulting from what is
  usually termed  ``the throat fitting
  into the bulk''   directly from the Klebanov-Strassler geometry.
  Generating
  the small value of the superpotential $W_{0}$ via the mechanism proposed by
  Demirtas, Kim, McAllister and Moritz (DKMM), we identify two
  possible DKMM-refined KKLT scenarios for stabilizing the light axio-dilaton modulus.
  The first scenario  spoils the expected hierarchy
  between the bulk and the throat mass scales and  implies
  that the energy scale of the uplift
  is larger than the species scale of the effective theory in the
  throat. Moreover it requires an unnaturally large tadpole $N\sim 10^{7-8}$
  that is certainly   in conflict with tadpole cancellation.
  For the less restricted second scenario,  the hierarchy can be
  controlled at the expense of, under the most optimistic assumptions,
  an only moderately large tadpole $N\sim 10^{2-3}$.
\end{abstract}

\clearpage


\section{Introduction}
\label{sec_intro}

The KKLT scenario \cite{Kachru:2003aw} has been proposed to be a controlled 
set-up of string moduli stabilization that eventually leads
to a dS minimum. It is important to see that it is not a fully fledged string construction
but rather consists of a number of ingredients that were argued to 
be generically present in consistent string models, working together
in an intricate manner to give dS vacua. In view of the swampland program \cite{Vafa:2005ui,Ooguri:2006in} (see \cite{Brennan:2017rbf,Palti:2019pca,vanBeest:2021lhn} for recent reviews) this construction has been under intense
scrutiny during the last years \cite{Obied:2018sgi,Ooguri:2018wrx,Garg:2018reu,Andriot:2018wzk,Bedroya:2019snp}, but we think it is fair to say that
the KKLT construction has nevertheless passed many non-trivial tests and
that it is still not really clear where it actually
fails.

Let us briefly review, how KKLT works.
One is working in an effective 4D supergravity theory that results 
from a compactification of the type IIB superstring with fluxes,
branes and orientifold planes. Then one proceeds in three steps:
First, stabilize  the complex structure and axio-dilaton moduli via
three-form fluxes in a no-scale non-supersymmetric Minkowski minimum with $|W_0|\ll 1$.
Second, one considers the effective theory of the light K\"ahler
    modulus  described by the leading order  K\"ahler potential and a superpotential
of the form $W=W_0 + A \exp(-aT)$.  
The non-perturbative effect conspires with the tiny value of $W_0$ to
stabilize $T$ in a supersymmetric AdS minimum.
Finally,  one uplifts the AdS minimum to dS via \mbox{$\overline{D3}$-branes} at the
    tip of a warped throat.

This set-up has been scrutinized from various sides in the past.
It was questioned  whether an 
$\overline{D3}$-brane at the tip of a
warped throat is really a stable configuration 
(see \cite{Danielsson:2018ztv} for a review).
Moreover, it has been
questioned whether the  4D description of the KKLT AdS minimum does
really uplift to a full   10D  solution of string theory
\cite{Moritz:2017xto,Kallosh:2018wme,
  Kallosh:2018psh,Gautason:2018gln,Hamada:2018qef,Hamada:2019ack,
  Carta:2019rhx,Gautason:2019jwq,Bena:2019mte,Blumenhagen:2020dea}.
More recently, it was pointed out in \cite{Gao:2020xqh} that a successful uplift
generically leads to a singularity, as   the warp factor in the vicinity
of where the non-perturbative effect is localized  becomes negative.
This was termed the singular-bulk problem.

There were also growing concerns even about step 1.
One needs $\vert W_0\vert \ll 1$, so that this should better  not be in the
swampland. In the large complex structure regime,
Demirtas, Kim, McAllister and Moritz (DKMM) provided
a mechanism  \cite{Demirtas:2019sip} that gives $W_0=0$ at leading order where
subleading (instanton-like) terms provide the stabilization of a
 final light complex structure modulus. This leads
to exponentially small values of $W_0$. For the later uplift
one actually needs a similar controllable mechanism  close to a
point in the complex structure moduli space  where one modulus
approaches a conifold point so that  large
warping can occur.  The generalization of the DKMM approach
to this so-called coni-LCS regime
has been performed in \cite{Demirtas:2020ffz,Alvarez-Garcia:2020pxd}.
There, also techniques for the determination of the periods of the
Calabi-Yau (CY) threefold in this regime of the complex structure moduli
space  have been developed (see also \cite{Alvarez-Garcia:2021mzv}).
A more detailed analysis using asymptotic
Hodge theory has been performed in \cite{Bastian:2021hpc}. One of their main
results is that, depending on the asymptotic Hodge structure,
an exponentially small $W_0$ does not necessarily mean that the scalar
potential leads also to exponentially small mass eigenvalues.
Other systematic studies of these so-called perturbatively flat flux
vacua for concrete
Calabi-Yau manifolds have been reported in \cite{Honma:2021klo,Demirtas:2021nlu,Demirtas:2021ote,Broeckel:2021uty,Carta:2021kpk,Carta:2022oex}.

Additionally, for the description of the uplift one  has to invoke an effective action that is
valid in the strongly warped regime. Based on earlier work
\cite{Douglas:2007tu}, 
this question has been addressed  recently 
\cite{Bena:2018fqc,Blumenhagen:2019qcg,Bena:2019sxm,Dudas:2019pls,Randall:2019ent}.
One of the main results of \cite{Bena:2018fqc} is that the uplift
term could destabilize\footnote{While this work was completed, we
  became aware of \cite{Lust:2022xoq}, in which this destabilization was
  questioned by computing more carefully the off-shell effective
  action for the conifold modulus $Z$. }
 the complex structure modulus $Z$, that controls the
size of the 3-cycle that shrinks to zero size at the conifold locus
$Z=0$. This provides a lower bound on the parameter $g_s M^2 \gtrsim
O(50)$ pointing into the direction that  KKLT is maybe  not compatible
with tadpole cancellation
\cite{Bena:2020xrh,Bena:2021wyr,Plauschinn:2021hkp,Grana:2022dfw},
which in fact is thought to be a genuine quantum gravity effect.

In \cite{Blumenhagen:2019qcg} the effective action in the
warped throat was analyzed in more detail, where the parametric dependence of all quantities
on the two moduli $\{Z,T\}$ and on the parameters  $\{g_s,M,N,y_{\rm UV}\}$  was
carefully determined.
Here $M$ is the flux at the bottom
of the throat, $N$ its D3-brane tadpole and $y_{\rm UV}$ the length of
the throat (in Klebanov-Strassler (KS) \cite{Klebanov:2000hb} coordinates) before it enters into the bulk Calabi-Yau.
It became evident  that it is the largish 
parameter $g_s M^2$ that controls the appearing mass hierarchies.
In particular,  the existence of a  red-shifted  tower of KK-modes was established, whose mass scale turned out to be close to  the
mass of the conifold modulus $Z$.
Since this
fitted nicely into the scheme of the
emergence proposal \cite{Heidenreich:2017sim,Grimm:2018ohb,Heidenreich:2018kpg,Corvilain:2018lgw}, it could not
conclusively be argued that the appearance of these ultra-light KK-modes signal
a complete breakdown of the effective theory\footnote{In \cite{Lust:2022xoq} these
  KK-modes were made responsible for the change in the off-shell action.}.

It is the purpose of this work to combine  the analysis of this latter
approach \cite{Blumenhagen:2019qcg,Blumenhagen:2020dea} with the recent results
\cite{Demirtas:2019sip,Demirtas:2020ffz,Alvarez-Garcia:2020pxd} about 
generating an exponentially  small $W_0$. We call this the ``DKMM-refined KKLT scenario''.
This is in the same spirit
as the analysis for KKLT-like AdS minima carried out recently in
\cite{Demirtas:2021nlu,Demirtas:2021ote},
where here we are less concerned about actual string model building on
concrete CY manifolds than on revealing the parametric dependence of
all involved mass scales. Generalizing the approach of 
\cite{Blumenhagen:2019qcg}, now  we
are careful about the three moduli $\{Z,T,S\}$ and the parameters
$\{M,N,y_{\rm UV},a\}$  where
$S$ denotes the axio-dilaton and $a$ the parameter in the
non-perturbative KKLT superpotential. 

The paper is organized as follows: In section 2 we review the local
Klebanov-Strassler geometry and derive  conditions for this
geometry to be glued  into an unwarped CY manifold. It turns out
that one of these conditions is the one that was termed ``the throat
fitting into the bulk'' in \cite{Carta:2019rhx}.
There a more indirect argument
based on the D3-brane backreaction was used, so that it is satisfying
that one can derive the same condition using the explicit flux supported KS solution. In section 3 we recall the effective theory
in the warped throat and the moduli stabilization scheme of 
\cite{Demirtas:2020ffz,Alvarez-Garcia:2020pxd} giving $\vert W_0\vert \ll 1$.
We also recall the appearance of ultra-light KK-modes and the relation
\mbox{to the emergence picture. We distinguish two possible scenarios
for the} stabilization of the light axio-dilaton modulus, labelled as
``Scenario 1'' and ``Scenario 2'' \mbox{in our paper. Scenario 1 seems to be more constrained than  Scenario 2, but} possibly only
because for the latter, we treat certain parameters as essentially arbitrary, even though
in concrete models they will also be determined by  three-form fluxes.

Section 4 deals with the final stabilization
of the K\"ahler modulus and the extra constraints arising from a
successful uplift to a dS minimum. We find a number of inequalities
the parameters need to satisfy.
For our more constrained Scenario 1, the constraints
turn out  to be not natural, in the sense that  exponentially
small numbers are  bounded from below by rational expressions
of the flux quanta. This has a number of consequences.

First, it turns out that the expected hierarchy between the bulk and
the throat mass scales is spoiled, invalidating the employed
low-energy effective action in the throat. In fact, the mass scale  of
the bulk complex structure moduli turn out to be smaller than the red-shifted
KK mass scale. This leads to a lowering of the (throat) species scale,
such that it becomes smaller than the 
energy scale of the uplift. We notice that in \cite{Lust:2022lfc} it was argued that
holography imposes a bound on the value of the cosmological constant
already in the supersymmetric AdS vacuum that leads to an analogous
inconsistency for the (bulk) species scale.
Second, the above mentioned constraints lead to 
much  larger tadpoles $N\sim 10^{7-8}$ than previously envisioned 
and are almost certainly in conflict with tadpole cancellation
respectively the recent tadpole conjecture
\cite{Bena:2020xrh,Bena:2021wyr,Plauschinn:2021hkp,Grana:2022dfw}.

For the less constrained  Scenario 2 the hierarchy could be controlled
by choosing still moderately large fluxes leading to a tadpole of the order $N\sim
10^{2-3}$. However, here certain parameters were treatad as freely
tunable, while in practice they will also depend on flux
quanta. Therefore, we expect the above tadpole to merely give a  lower
bound that in reality will easily be exceeded by orders of magnitude.

As proof of principle, we provide in section 5 explicit examples 
showing the analytically derived behaviour for the effective theory in
the throat. Putting all evidence  together, we would like
to interprete these results as an indication that after all, the 
DKMM-refined  KKLT construction might turn out to be   inconsistent with quantum
gravity, hence in the swampland. 
We note that a recent  analysis \cite{Junghans:2022exo,Gao:2022uop,Junghans:2022kxg} of the
regions of control in the context of the Large Volume Scenario
arrived at a very similar conclusion.


\section{Geometric constraints in the  warped throat}
\label{sec_volbound}

In this section, we  derive a couple
of geometric constraints arising from designing a controlled picture
of a  Calabi-Yau compactification that develops a strongly warped throat
close to a conifold point in the complex structure moduli space.
These constraints will arise from gluing the local Klebanov-Strassler
solution to a bulk unwarped CY manifold.
In particular, we  directly derive the bound for the Calabi-Yau volume modulus
resulting from ``the throat fitting into the bulk'' in \cite{Carta:2019rhx}.

Note that throughout this paper, we will only explicitly determine
the parametric dependence suppressing numerical prefactors.
Moreover, though certainly present we will not explicitly
consider the compact axionic partners of the three moduli, i.e.
$\{\arg(Z),{\rm Im}(T),{\rm  Im}(S)\}$,
which are stabilized by the same fluxes and non-perturbative
effects that stabilize the saxions. Moreover, the masses of these
axions are expected to scale in the same way as those of their saxionic partners.
We are using natural units, i.e. $M_{\rm pl}=1$ except when it is
explicitly written.

\subsection{The strongly warped regime}

Our starting point is type IIB string theory compactified on a CY
orientifold, where three-form fluxes will stabilize some of the complex structure  moduli
and the axio-dilaton.
Strongly warped throats can develop close to a conifold point $Z=0$ in the complex structure moduli space. 
There exists a proposal for the effective action of the conifold
modulus $Z$ and the warped volume modulus ${\cal V}_w$ in the
strongly warped regime \cite{Douglas:2007tu, Bena:2018fqc}. This is supposed to be valid
at a point very close to a conifold singularity where warping effects
become relevant.
At such points the geometry develops a long throat,
at the tip of which a three-cycle $A$ becomes  small. 
The corresponding complex structure modulus is defined as $Z=\int_A
\Omega_3$.

Such a warped throat can be supported
by turning on R-R three-form flux $M$ on the A-cycle 
and an NS-NS three-form flux $K$ on the dual B-cycle.
This induces a contribution to the D3-brane tadpole
\eq{
  \label{D3tadpole}
  N=M\cdot K\,
}
and   self-consistently fixes the conifold modulus
at the exponentially small value
\eq{
  \label{conifoldmod}
          |Z|\sim\exp\left( -{2\pi K\over g_s M}\right)
}
for $N\gg g_s M^2$. The full ten-dimensional metric can be parameterized as \cite{Giddings:2001yu} 
\begin{equation} 
    ds^2 = e^{2A(y)} g_{\mu \nu} dx^{\mu} dx^{\nu} + e^{-2A(y)} \tilde{g}_{mn} dy^{m} dy^{n}\,
\end{equation}
where $g_{\mu \nu}$ is the four-dimensional spacetime metric and $\tilde{g}_{mn}$ the Ricci-flat metric of the CY with coordinates $y^{m}$. The warp factor $A(y)$ depends only on these internal coordinates. The regime of strong warping is given by
\eq{
    \label{strongwarping}
    \mathcal{V}_{w} \abs{Z}^2 \ll 1
}
where the warped volume  $\mathcal{V}_{w}$ (measured in units of $\alpha'$) of the CY is given by
\begin{equation}
  \label{eq:warpvol}
   \mathcal{V}_{w} = \frac{1}{g_{s}^{3/2}(\alpha')^3}\int_{\rm CY_{3}} d^6 y \, e^{-4A(y)} \sqrt{\abs{\tilde{g}}} \sim \tau^{\frac{3}{2}}\,.
\end{equation}
The string coupling $g_{s}$ is related to the axio-dilaton $S =e^{-\phi} + iC_0$ via
$g_{s} = \langle s\rangle ^{-1}$, with the saxion $s=e^{-\phi}$.  The
saxion $\tau$ appears in the complexified K\"ahler modulus $T = \tau + i\theta$.

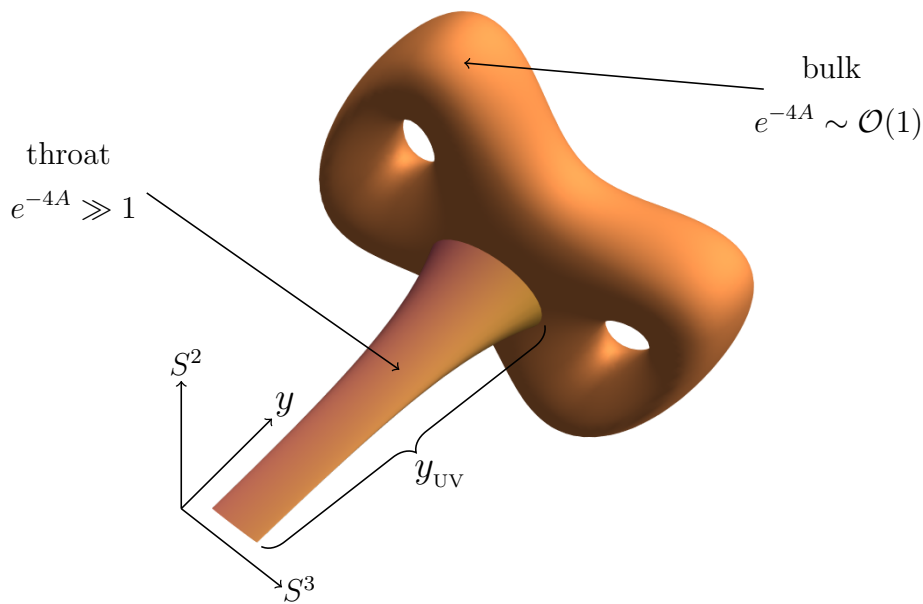
\begin{figure}[ht]
\centering
\begin{tikzpicture}
\node[anchor=south west, inner sep=0] (sketch) at (0,0) {\includegraphics[width=0.55 \textwidth]{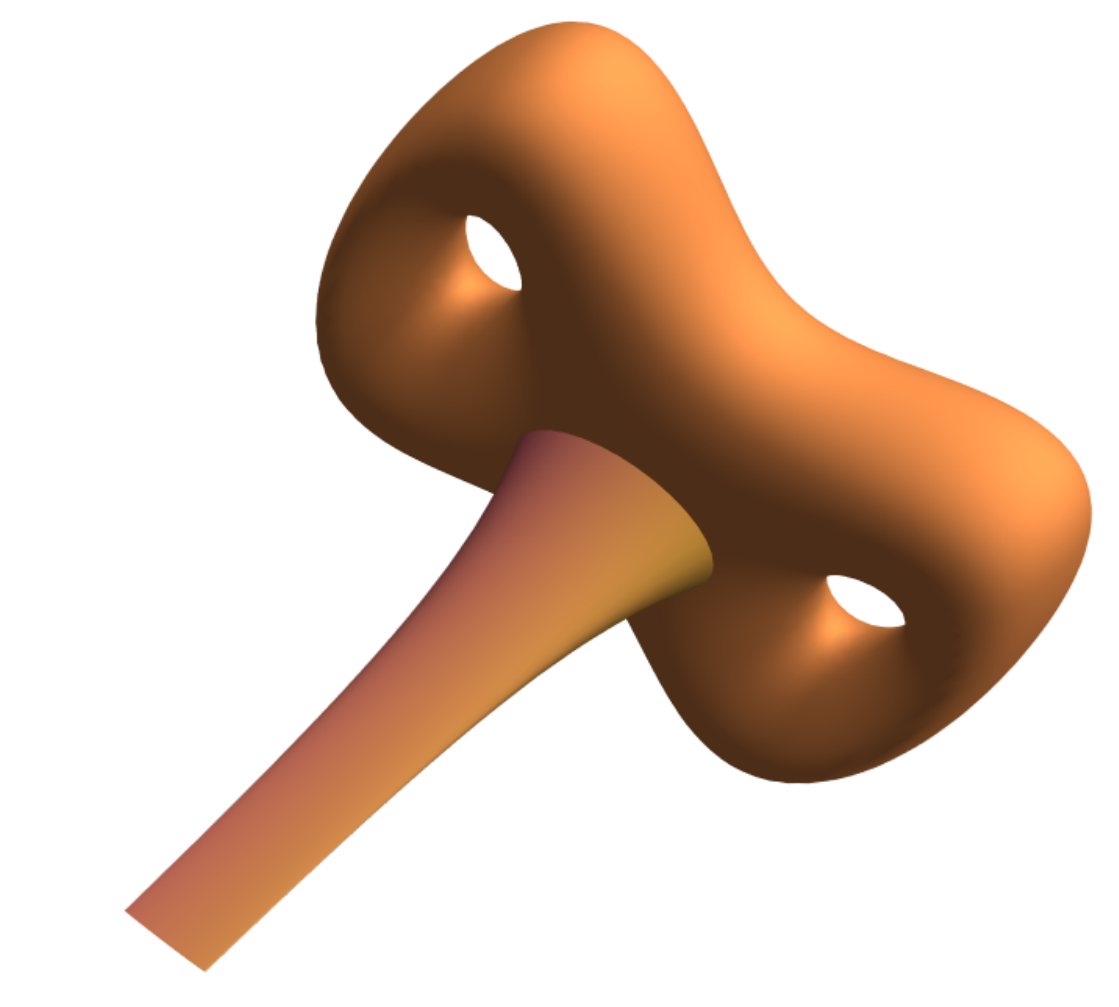}};

\draw[->,semithick]  (53.97:.88) -- ( $  (53.97:.88) + (-38 : 1.7) $ );
\node(S3) at ($ (53.97:.88) + (-34 : 1.9) $) {$S^3$};

\draw[->,semithick]   (53.97:.88) -- ( $   (53.97:.88)+ (44.80 : 1.7) $ );
\node(Y) at ($  (53.97:.88)+ (45 : 1.95) $) {\large $y$};

\draw[->,semithick]  (53.97:.88) -- ( $ (53.97:.88)+ (90 : 1.7) $ );
\node(Y) at ($  (53.97:.88)+ (88 : 1.95) $) {$S^2$};

\draw [decorate,decoration={brace,mirror, amplitude=8pt},semithick,xshift=-3pt,yshift=0pt]
		( 7:1.7) -- (30:6.3) node [black,midway,xshift=.5cm,yshift=-.5cm] {\large  $y_{_\text{UV}}$};

\draw[->,semithick] ($(57.22: 7.9)+(-5:4)$) -- (57.22: 7.9);
\draw[->,semithick]  ($(36.84:4.27)+(145:4.1)$) -- (36.84:4.27) ;

\node(bulk) at ($(57.22: 7.9)+(-5:5)$) {$\begin{aligned}\text{bulk \,\,\,\,\,\,\,\,\,\,}\\e^{-4A} \sim   \mathcal{O} (1) \end{aligned}$};  
\node(throat) at ($(36.84:4.27)+(150:5)$) {$\begin{aligned}\text{throat \,\,\,}\\e^{-4A} \gg  1  \end{aligned}$};

\end{tikzpicture}
\caption{A schematic figure of a Calabi-Yau with a KS-throat. The KS-throat is glued to the bulk at the UV scale $y_{\rm UV}$.} \label{KS_solution}
\end{figure}

A warped geometry on the deformed conifold is locally described by the
Klebanov-Strassler (KS) solution \cite{Klebanov:2000hb}. This is a
cone over $T^{1,1}$, which is cut off in the IR by a finite size
$S^3$. The metric of the KS-throat is explicitly known 
\eq{
\label{KSmetric}
  \widetilde{ds}^2_{\rm KS} = \frac{1}{2} \abs{S}^{\frac{2}{3}} K(y) \Big[ \frac{dy^2+ (g^5)^2}{3K^3(y)} + &\cosh^2{\Big(\frac{y}{2}\Big)}((g^3)^2+(g^4)^2) \\
     +&\sinh^2{\Big(\frac{y}{2}\Big)}((g^1)^2+(g^2)^2) \Big]\,
   }   
where the collection of einbeins $\{g^{i}\}$ provides a basis for the
$S^{2} \times S^{3}$ base of the cone.
As explained in \cite{Blumenhagen:2019qcg}, the deformation parameter $S$ is related to the conifold modulus $Z$ by a rescaling
\begin{equation}
    S = (\alpha')^{\frac{3}{2}} \sqrt{g_{s}^{\frac{3}{2}}\mathcal{V}_{w}} Z\,
\end{equation}
and has units of $[\text{length}]^3$. The factor $K(y)$ is given by
\begin{equation}
    K(y) = \frac{\left(\sinh{(2y)} - 2y\right)^{\frac{1}{3}}}{2^{\frac{1}{3}} \sinh{(y)}}\,
\end{equation}
and at leading order the warp factor takes the form \cite{Klebanov:2000hb,Herzog:2001xk}
\begin{equation}
  \label{warpKS}
    e^{-4A(y)} \approx  1+2^{\frac{2}{3}}\frac{(\alpha' g_{s} M)^2}{\abs{S}^{\frac{4}{3}}} \mathcal{I}(y) = 1+2^{\frac{2}{3}} \frac{g_{s} M^2}{(\mathcal{V}_{w} \abs{Z}^2)^\frac{2}{3}} \mathcal{I}(y)
\end{equation}
with
\begin{equation} 
    \mathcal{I}(y) = \int_{y}^{\infty} dx \frac{x \coth(x)-1}{\sinh^2(x)} (\sinh(2x) - 2x)^{\frac{1}{3}}\,.
\end{equation}
In the following, as usual we think of the total geometry as such a KS-throat of  length
$y_{\rm UV}$ glued to the unwarped bulk Calabi-Yau manifold.
A sketch of the CY is shown in figure \ref{KS_solution}.

\subsection{Geometric constraints}

Generalizing the results from \cite{Blumenhagen:2019qcg},
let us now derive a couple of  constraints on the parameters $\{g_s,M,N,y_{\rm UV}\}$ and
the moduli $\{\tau,Z\}$ that guarantee a parametrically controlled geometry of this type.
In particular we  will derive  bounds on  the allowed length $y_{\rm
  UV}$ of the KS throat which upon combination will lead to 
a lower bound for the volume modulus $\tau$ in terms of the D3 tadpole
contribution $N = MK$.

First, in order for the supergravity, large radius description to be consistent
one demands that the  size of the $S^3$ at the tip of the
throat  stays larger than the string length.  This can be 
read-off  from the KS metric  \eqref{KSmetric} with the warp factor \eqref{warpKS}
\eq{
       R^2_{S^3}\sim e^{-2A(0)} \alpha'\, g_s^{1/2} \,({\cal V}_w
       |Z|^2)^{1\over 3} \sim \alpha' g_s \vert M \vert 
}
leading to the constraint
\eq{
\label{sugraregime}
                   \boxed{      g_s \vert M \vert > 1}\,.
 }
                       
Second, in the region where the throat is glued to the  bulk CY,
warping should be negligible
meaning that the warp factor should be of $\mathcal{O}(1)$. By using
the leading order behaviour of $\mathcal{I}(y)$ for large $y$
we end up with the lower bound
\eq{
  \label{eq:ylowerbound}
                   e^{{4\over 3} y_{\rm UV}}>\left({ g_s M^2\,  y_{\rm UV}
                     \over \tau }\right) {1 \over |Z|^{4\over 3}}\,.
 }

Third, we require that the throat fits into the bulk, which means
that the contribution of the throat  to the warped volume \eqref{eq:warpvol}
must be smaller than the value of the total volume  ${\cal V}_w$. To obtain the volume of the KS throat, we proceed as in formula
\eqref{eq:warpvol} , but now we integrate the radial coordinate over
the restricted domain $y \in [0,y_{\rm UV}]$. Making use of the
diagonal basis for the KS metric and the expression for the warp factor we find the relevant scaling to be
\begin{equation}
  \label{hannover96}
    \mathcal{V}_{w}^{\rm  KS} \sim g_{s} M^2\,
    (\mathcal{V}_{w} \abs{Z}^2)^\frac{1}{3} \int_{0}^{y_{\rm UV}} dy \: \mathcal{I}(y) \sinh^2(y)\,.
\end{equation}
For large $y_{\rm UV}$ the integral $\mathcal{I}(y)$ found on the right-hand side behaves like
\begin{equation} 
    \mathcal{I}(y) \sim y\, e^{-4y/3}
\end{equation}
so that the integral in \eqref{hannover96} can be approximated by
\begin{equation}
     \int_{0}^{y_{\rm UV}} dy \: \mathcal{I}(y) \sinh^2(y) \approx
     y_{\rm UV}\,
     e^{2y_{\rm UV}/3}\,.
\end{equation}
Thus, imposing  $\mathcal{V}_{w}^{\rm  KS}<\mathcal{V}_{w}$ leads to the upper bound
\eq{
  \label{eq:yupperbound}
        e^{ {4\over 3}y_{\rm UV}}<  \left( {\tau\over g_s M^2\,  y_{\rm
            UV} } \right)^{2} {1 \over |Z|^{4\over 3}} \,.
}
Consistency of  this with the lower bound \eqref{eq:ylowerbound}
implies the condition
\eq{
  \label{reltaua}
  \tau > g_s M^2\,  y_{\rm UV}\,.
}
Noticing that the scaling of the lower \eqref{eq:ylowerbound} and the upper
\eqref{eq:yupperbound} bound with $|Z|$ is the same and that
according  to \eqref{conifoldmod} $|Z|^{-4/3}$ is the
only exponentially large quantity on the right hand side of the
bounds, the geometric cut-off $y_{\rm UV}$ must scale
like
\eq{
  \label{valueofyuv}
         y_{\rm UV}\sim -\log |Z| \sim  {2\pi N\over g_s M^2}\,.
}     
For $N\gg  g_s M^2$ this is indeed a large number, i.e. our initial assumption $y\gg 1$ was
justified. Combining the two conditions \eqref{reltaua}  and
\eqref{valueofyuv}  yields the intriguing relation
\eq{
\label{werder}
\boxed{ \phantom{\Big(}  \tau >2\pi N\ \, }\,,
}
i.e. the ``size'' of the CY is larger than the D3-brane tadpole. We notice that 
this is the same condition as proposed in equation (5.7) of \cite{Carta:2019rhx} via a more
indirect argument based on the D3-brane backreacted geometry.
It is very satisfying that the same condition also arises
as a geometric constraint for the explicit KS metric of the deformed
conifold.


\section{Mass hierarchies}
\label{sec_moduli}

Recall that the first step in the KKLT construction is to stabilize the complex
structure moduli such that the value of the superpotential
in the minimum is exponentially small, i.e. $\vert W_0\vert \ll 1$.
How this can be achieved in the large complex structure regime
was described by Demirtas, Kim, McAllister and Moritz in \cite{Demirtas:2019sip}.
The generalization of this algorithm to the conifold regime
and the necessary computation of the periods in this regime
was done in \cite{Demirtas:2020ffz,Alvarez-Garcia:2020pxd}. For more details we refer the reader
to these papers.

Here we take these results  and analyze what the implications
for the various mass scales in the KKLT scenario are.
The algorithm allows us to find concrete flux configurations which yield a perturbatively flat direction of the vacuum satisfying
\begin{equation}
    W_{\rm pert}\rvert_{\vec{U}} = 0\,,\qquad {\rm and}\qquad \vec{U} = \vec{p} \,S
\end{equation}
with $\vec{p}$ being related to the flux vectors. This means that all
complex structure moduli are fixed in terms of the axio-dilaton.
The masses of the massive complex structure moduli scale as
  \eq{
    \label{masscomplexstr}
                                 M_{\rm cs}\sim {g_s^{1\over 2}\over
                                   {\cal V}_w} M_{\rm pl}\,.
}
Since the string and the bulk Kaluza-Klein scales are actually
$M_{\rm s}\sim   g_s^{1/4}M_{\rm pl}/{\cal V}^{1/2}_w $ and  ${M_{\rm KK}\sim   M_{\rm pl}/{\cal V}^{2/3}_w}$, $M_{\rm cs}$ is the lightest mass scale
in the bulk\footnote{We will denote bulk masses by capital letters and throat masses by lowercase letters.}.

After integrating out the $U_i$ one gets an effective superpotential
for the remaining two moduli, $Z$ and $S$
\eq{
\label{superpotb}
     W=-{M\over 2\pi i} Z \log Z  +i  {K S} Z  + M_1 Z+O(Z^2,e^{-S})\,,
}
where the non-perturbative $\exp(-S)$ terms arise from the world-sheet instanton corrections
$\exp(-U_i)$ (in the mirror dual picture). The coefficient $M_1$ is a complex parameter derived from the flux vectors \cite{Alvarez-Garcia:2020pxd}. Note that in the second
step of the KKLT construction
also non-perturbative terms in the K\"ahler modulus $T$ will be taken
into account.
Since for KKLT we need to work in the strongly warped throat, we
employ the corresponding K\"ahler potential 
from \cite{Douglas:2007tu, Douglas:2008jx}\footnote{The approximate
  numerical value of the order one constant $c'$ was
  determined in \cite{Douglas:2007tu}
  to be $c'\approx 1.18$. } 
\eq{ 
\label{kaehlerpotb}
    K=-3\log(T+\ov T) -\log(S+\ov S)+  {2  c'   M^2 \vert Z\vert^{2\over
      3}\over (S+\ov S)(T+\ov T)} + O(\xi^2)\,
}
where we have promoted the string coupling $g_s$ to the full
complex field $S$.
There will be higher order corrections in $\xi\sim   M^2 \vert Z\vert^{2\over
  3}/(S+\ov S)(T+\ov T)$ which are under control if
\eq{
  \label{controlcond}
  \left( |Z|\over  {\cal V}_w \right)^{2\over 3}\lesssim {1\over g_s M^2}\,.
}
This K\"ahler potential and the superpotential \eqref{superpotb}
are the defining data of the effective strongly warped KKLT model
whose mass scales  we analyze in more detail in the following.
The resulting scalar potential takes the usual form
\eq{
  V=e^K \left( G^{I\ov J} D_I W D_{\ov J} \ov W -3 |W|^2
  \right)\,
}  
with $I,J\in\{Z,S,T\}$.

\subsection{Conifold modulus and axio-dilaton}
\label{Conifold modulus and axio-dilaton}

Next one stabilizes the conifold modulus. As shown in \cite{Blumenhagen:2019qcg},
the leading order kinetic term for the conifold modulus scales
just in the right way to admit a deformed no-scale structure for the
K\"ahler modulus.
Since there the explicit dependence on the field $S$ was
not taken into account, let us extend this deformed  no-scale
structure  to the three moduli case.
Using the K\"ahler potential \eqref{kaehlerpotb}, one still finds some leading order
cancellations 
\eq{
       G^{S\ov T}&=0 \\
      G^{Z\ov A} \partial_{\ov A}K &= 0 + O(\xi) \\
      G^{A\ov B} \partial_A K \partial_{\ov B}K &= 3+ O(\xi)\\      
      G^{S\ov Z} \partial_{\ov Z}K  &=0 + O(\xi)
}      
with the sums over the $A,B$ indices restricted to  $A,B\in\{Z,T\}$.
Using these relations, it follows that for a superpotential $W(Z,S)$
the scalar potential is at leading order given by
\eq{
  V\approx e^K \left( G^{M\ov N} {\cal D}_M W {\cal D}_{\ov N} \ov W 
  \right)\,,\qquad M,N\in\{Z,S\}
}
with
\eq{
         {\cal D}_Z W := \partial _Z W\,,\quad  {\cal D}_S W := D_S
         W=\partial_S W +\partial_SK\, W\,.
}
Hence, $\partial_Z W=0$ fixes the conifold modulus at the exponentially small value 
\eq{\label{Z_minimum}
Z\sim \zeta_0 \exp\left(-{2\pi K\over  M} S\right)\,, \qquad
\text{with }\ \zeta_0=e^{2\pi i \frac{M_1}{M}-1}\,.
}
Plugging this back into the superpotential \eqref{superpotb} leads to an effective
superpotential for the stabilization of $S$
\eq{
  \label{racetrack}
        W_{\rm eff}={1\over 2\pi i} MZ + O(Z^2,e^{-S})\sim  \frac{M}{2\pi i}\zeta_0\, e^{-{2\pi
            K\over  M} S}+  a_{1}\, e^{-c_{1} S} + a_{2}\, e^{-c_{2} S}+\ldots
      }
where we have indicated the first two leading (mirror dual) world-sheet
instanton corrections.
Hence, the stabilization of the axio-dilaton can occur via a race-track
scenario (see \cite{Kallosh:2004yh} for a KKLT application).
Recall that for large $s$ one can approximate the solution to the
F-term equation for $S$ by solving $\del_{S} W = 0$ instead of $D_S
W=0$. This yields
\begin{equation}
  \label{smin}
    S_{0} = \frac{1}{c_{1} - c_{2}} \log \left(- \frac{c_{1} a_{1}}{c_{2} a_{2}} \right)
\end{equation}
so that  $\abs{c_{1} - c_{2}} \ll c_2$ is necessary to get large
$s_0=g_s^{-1}$.
The value of the superpotential  at the minimum can be approximated by
\eq{\label{approximate_W0}
    W_{0} \sim {c_{2}-c_{1} \over c_{2}} a_{1} e^{-c_{1} S_{0}}\,.
    }
Thus,  for a successful controllable stabilization of the axio-dilaton, 
the involved coefficients $(M,K,M_1,a_{1}, a_{2}, c_{1},c_{2})$ have to be tuned to a
certain degree. In practice these parameters are determined by the
underlying CY manifold and the fluxes turned on
\cite{Demirtas:2019sip,Demirtas:2020ffz,Alvarez-Garcia:2020pxd}. 
For our purpose, let us distinguish two  possible scenarios:
\begin{description}
    \item {\bf Scenario 1:} the $MZ$-term contributes dominantly to the race-track potential so
    that stabilization requires $c_{1}\approx 2\pi K/M$ and one finds
    \eq{
        |W_0|\approx {g_{s} M^2\over \left(2
            \pi \right)^2|K|} |Z|\,.}
    \item {\bf Scenario 2:} the $MZ$-term is sub-leading so that the two leading instanton corrections stabilize $S$. In this case, we can at least say
    \eq{
        |W_0| > {g_{s} M^2\over \left(2
            \pi\right)^2|K|} |Z|\,.}
\end{description}      

Next, let us estimate the masses of the two moduli $Z$ and $S$ which are given by the eigenvalues of the mass matrix
 \begin{equation}
   \label{eq:massmatrix}
    M^{i}_{\,\,\,k} = G^{ij} \partial_{j} \partial_{k} V\,
  \end{equation}
where $G^{ij}$ denotes the (inverse) K\"ahler metric and $V$ is the full scalar potential of our theory. 
Here the transformation to a canonically normalized basis is already taken care of.
The mass of the conifold modulus turns out to be 
\eq{
\label{massconi}
                m_Z\simeq  {1\over  (g_s M^2)^{1\over 2}} \left(
                  |Z|\over  {\cal V}_w \right)^{1\over 3}  M_{\rm pl} \,
}
and for the second and the first scenario, the mass of $S$ can be expressed as\footnote{As pointed out in  \cite{Bastian:2021hpc}, at the coni-LCS boundary
it can in principle  happen that despite the exponentially small value
of $W_0$, the modulus $S$ receives a polynomial mass.}
\eq{
    \label{massdilaton}
    m^{(2)}_S\simeq  { c_1\, c_2 \, |W_0|\over  g_s^{3\over 2}  {\cal V}_w}   M_{\rm pl}\,,\qquad \quad  m^{(1)}_S\simeq {N  \over
    (g_s M^2)^{1\over 2}} \left(
    |Z|\over  {\cal V}_w \right) M_{\rm pl}\,.
}   
In Scenario 1, we find the ratio
\begin{equation}
{ m^{(1)}_S  \over m_Z } = {N \over \tau} |Z|^{2\over 3} \ll 1\,,
\end{equation}
where we have used that $\tau > N$ in the last step. The $S$ modulus
is therefore much lighter than the $Z$ modulus in Scenario 1, just as
assumed before. Later in section \ref{Mass hierarchies extended} we
will derive a sufficient condition to have the desired hierarchy $m^{(2)}_S < m_Z$ in Scenario 2.

However, let us already mention that with respect to $|W_0|$ the mass
of the dilaton scales in the same
way as the mass of the  K\"ahler modulus, to be discussed
next. Therefore we also stabilized
them simultaneously in our actual computations. To leading order, we
got the same masses for the moduli $S$ and $T$ and the same minimum
conditions as in the step-by-step procedure presented
for pedagogical reasons in this paper.

\subsubsection*{Destabilization of conifold modulus}

For completeness, let us mention an issue that was first observed in
\cite{Bena:2018fqc} (and questioned recently in \cite{Lust:2022xoq}).
Setting $S$ to its VEV and
adding the contribution of an  $\ov{D3}$-brane  at the tip of the
throat to the scalar potential one obtains
\eq{
\label{Vantitot}
V_{\rm tot}= {9\over 8 c' M^2} {|Z|^{4\over 3}\over {\rm Re}(T)^2} \bigg[
      \Big\vert{\textstyle-{M\over 2\pi i}\log Z+ i{K\over g_s}+ M_1 -{M\over 2\pi i}}\Big\vert^2 +
      {c' c''\over g_s} \bigg]\,.
}
Thus, both the three-form flux and the $\ov{D3}$-brane contribution
scale in the same way with the moduli $Z$ and $T$.
Plugging in concrete numbers\footnote{The order one constant
  $c''=2^{1\over 3}/\mathcal{I}(0)\approx 1.75$ as
  mentioned in \cite{Bena:2018fqc}.}  for the constants
$c',c''$, it was shown 
that a minimum in the $Z$ coordinate only continues to exist  for
$\sqrt{g_s} \vert M\vert > 6.8$.

\subsection{Throat KK-modes and emergence}

In the strongly warped throat
geometry one direction (namely $y$ along the throat) becomes
much larger than the other bulk directions. Therefore, one is
dealing with a highly non-isotropic geometry that could support
very light Kaluza-Klein modes \cite{Frey:2006wv, Burgess:2006mn,Shiu:2008ry,deAlwis:2016cty}.
It was explicitly shown in  \cite{Blumenhagen:2019qcg}  that such modes indeed
exist and that, reminiscent  of infinite distance limits their mass scale is
related to the distance of the conifold point in the complex structure
moduli space via emergence. This picture allowed one to derive
the cut-off scale $\Lambda$ of the effective theory in the throat.
Since there are KK-modes lighter than this cut-off, the actual
cut-off is the species scale \cite{Dvali:2007hz, Dvali:2007wp} $\tilde \Lambda_{\text{sp}}=\Lambda/\sqrt{N_{\rm sp}}$\,\,, where $N_{\text{sp}}$ denotes the number of light species below the species scale.

\begin{figure}[ht]
\centering
\includegraphics[scale = 0.44]{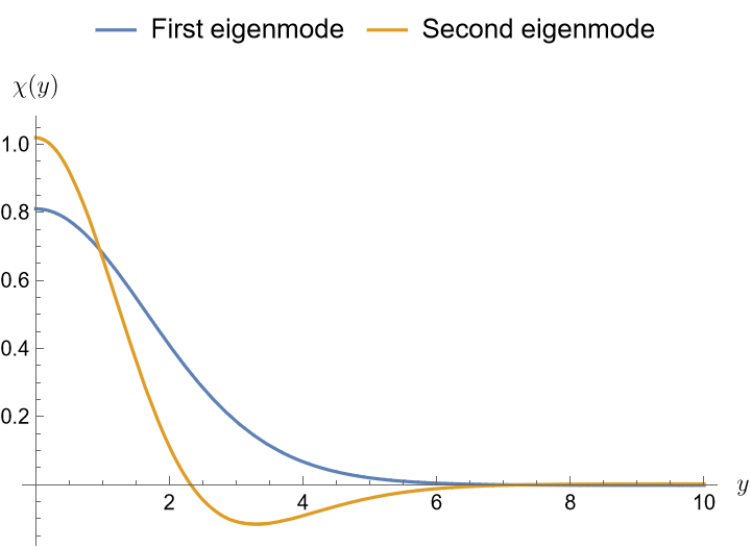}
\caption{Graphs of the numerically determined eigenfunctions associated
  to the two lowest modes, assuming $y_{\rm UV} = 10$.} 
\label{KK_eigenmodes_numerical}
\end{figure}

Following \cite{Blumenhagen:2019qcg}, we numerically
determined the masses of the KK-modes in a long warped throat, i.e. $y_{\rm UV} \gg 1$. For that purpose, we solved the same six-dimensional warped Laplace equation for the radial part of the eigenmodes. The equation reads
\begin{equation} 
    3K^2(y) \partial_{y}^2 \chi(y) + 4 \frac{\partial_{y} \chi(y)}{\sinh(y) K(y)} + k^2 \mathcal{I}(y) \chi(y) = 0
\end{equation}
where 
\begin{equation}
    k^2 = \frac{\alpha' g_{s}^\frac{3}{2} M^2}{2^\frac{1}{3} (\mathcal{V}_{w} \abs{Z}^2)^\frac{1}{3}} m_{\text{KK}}^2
\end{equation}
depends on the Kaluza-Klein masses. By respecting the Neumann boundary conditions
\begin{equation}
    \partial_{y} \chi_{\rm n}(y) \rvert_{y = 0} = 0 \quad \text{and} \quad
    \partial_{y} \chi_{\rm n}(y) \rvert_{y = y_{\rm UV}} = 0
\end{equation}
and the normalization requirement 
\begin{equation}
    \int_{0}^{y_{\rm UV}} \abs{\chi_{\rm n}(y)}^2 = 1 
\end{equation}
for all  mode numbers ${\rm n}$, we find eigenfunctions which are localized near
small values of $y$. The plots of the lowest two eigenmodes are shown
in figure \ref{KK_eigenmodes_numerical}. 

\begin{figure}[ht]
\centering
\includegraphics[scale = 0.59]{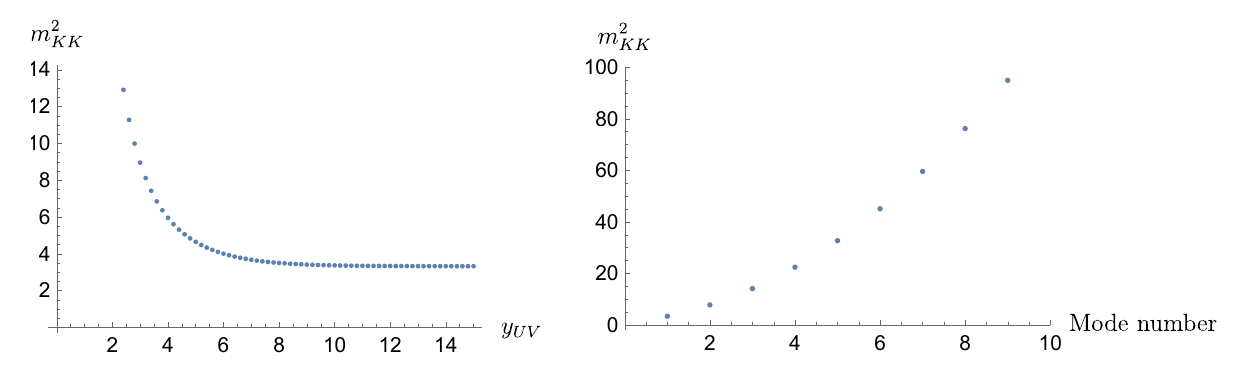}
\caption{Left: KK mass squared of the first eigenmode as a function of
  $y_{\rm UV}$. Right: KK mass squared of the nine lowest eigenmodes,
  assuming $y_{\rm UV} = 10$.} \label{function_mkk}
\end{figure}

The masses scale
approximately linearly with the mode number. The right plot in figure
\ref{function_mkk} shows this behaviour for the first few mode numbers
for fixed $y_{\rm UV}$.
Remarkably, for large values  $y_{\rm UV}>y^*_{\rm UV}$ the masses become
independent of the UV-cutoff. This is reflecting that  the
eigenfunctions are localized
near small $y$ so that the eigenvalues reach an asymptotic value beyond
$y^*_{\rm UV} \approx 10$.
In the left-hand side of figure \ref{function_mkk} we display this behaviour for the first eigenmode. 

To summarize, we obtain the following scaling of the red-shifted KK masses
\begin{equation}
    \label{m_KK}
    m_{{\rm KK, n}}^2 \sim \frac{(\mathcal{V}_{w} \abs{Z}^2)^\frac{1}{3}}{g_{s}^\frac{3}{2} M^2} {\rm n}^2 M_{\rm s}^2 \sim \frac{1}{g_{s}M^2} \left( \frac{\abs{Z}}{\mathcal{V}_{w}} \right)^\frac{2}{3} {\rm n}^2 M_{\rm pl}^2\,.
\end{equation}
Comparison with the mass of the $Z$-modulus from the previous subsection yields
\begin{equation}
      \frac{m_{\rm KK,n}}{m_{Z}}  = c\, {\rm n} 
\end{equation}
where in the regime $y_{\rm UV}< y^*_{\rm UV}$ the coefficient $c$
also contains a factor $y_{\rm UV}^{-1}$, which we are not explicitly
writing in the following. The vast majority of KK-modes is
therefore expected to be heavier than the conifold modulus.
Thus, the KK-modes  are parametrically of the same mass as the conifold modulus $Z$ and the description in terms of the effective action is at its limit.

As shown in \cite{Blumenhagen:2019qcg},
the existence and mass scale of these red-shifted KK-modes is consistent with the picture of
emergence of the field space metric \cite{Heidenreich:2017sim,Grimm:2018ohb,Heidenreich:2018kpg,Corvilain:2018lgw},
which means that integrating them out  the field
space metric should get a one-loop correction proportional to the tree-level
result. Assuming a one-dimensional tower of red-shifted KK modes,
the one-loop correction came out as
\begin{equation}
\label{eq:KK1loop}
	\begin{aligned}
		 g_{Z\ov Z}^{\rm 1-loop}
        \sim N_{\rm sp}^3\frac{1}{g_s M^2 }\frac{1}{\left(\mathcal{V}_w|Z|^2\right)^{2/3}}\,,
	\end{aligned}
\end{equation}
so that matching this  with  the tree-level metric resulting from \eqref{kaehlerpotb} fixed the number $N_{\rm sp}$ of light species as 
\begin{equation}
	N_{\rm sp}\sim\left(g_s M^2 \right)^{2/3}\,.
      \end{equation}
Using this scaling, we found that the one-loop corrections $g_{I\ov J}^{\rm 1-loop}$ of the field space metric, where $I,J\in\{Z,S,T\}$, are proportional to the tree-level expressions of the field space metric coming from the K\"ahler potential
$K\sim M^2\vert Z\vert^\frac{2}{3}/ (S+\overline{S}) (T+\overline{T})$.

From the spacing and number of modes below the cutoff, one could
derive  for the  UV cut-off
\eq{   \label{cut-off_Lambda}
    \Lambda\sim \sqrt{g_s M^2} \left({|Z|\over {\cal
          V}_w}\right)^{\!{1\over 3}} M_{\rm pl}\,  
  }
which intriguingly equals the mass of a D3-brane wrapping the A-cycle $S^3$
which vanishes at the conifold singularity.
Then, for the generalized species scale we get
\eq{ \label{species_scale}
  \tilde \Lambda_{\text{sp}}={\Lambda\over \sqrt{N_{\rm sp}}} \sim {\Lambda\over \left(g_s M^2 \right)^{1/3}}\,.
}
We can express $m_Z\sim \Lambda/(g_s M^2)$ so that in the regime of
interest, $g_s M^2 \gg 1$, one has the expected hierarchy  $m_Z<\tilde
\Lambda_{\text{sp}}$. 

In summary, up to this point we have the following hierarchy of mass
scales in  the warped throat
\eq{
  \label{masshierarchya}
     \Lambda >  \tilde \Lambda_{\text{sp}} > m_Z\sim m_{\rm KK} \,.
}
How the mass scales of the light $S$ modulus \eqref{massdilaton} and the K\"ahler modulus
fit into this hierarchy will be discussed in the next section. The
bulk masses $M_{\rm cs}<M_{\rm KK}<M_{\rm s}$ are assumed to be much heavier.


\section{K\"ahler modulus}

Now, let us consider the second step  of KKLT and, after integrating
out $Z$ and $S$, consider the effective superpotential
\eq{
  \label{Wkaehler}
  W_T=W_0 +A\exp(-aT)
}
where $W_0$ is the exponentially small value from the previous race-track
model for the $S$ modulus.
The parameter  $A$ is the so-called 1-loop Pfaffian, considered to be
independent of the complex structure moduli.
The parameter $a$ is defined as
\eq{
  a={2\pi\over N_c} \gamma
 }
where $N_c$ is the rank of the gauge theory featuring gaugino
condensation
and $\gamma$ is related to the size of the bulk 4-cycle $\tau_{\rm np}=\gamma\tau$
supporting this gauge theory. Note that for  an isotropic bulk Calabi-Yau
manifold  one expects $\gamma=O(1)$ and that an anisotropic CY
requires $h_{11}>1$.

One comment might be in order here. While this work was in preparation,
  employing holography, the authors of \cite{Lust:2022lfc} suggested that 
  no supersymmetric AdS minimum \`a la KKLT can possibly exist in a
  controlled  manner.  In particular, they argue that the AdS energy scale is larger than
  the (bulk) species scale, leading at best to a non-scale-separated
  AdS minimum. In addition, for the DKMM construction of perturbatively flat flux vacua,
  the authors argue that no supersymmetric AdS minimum can potentially
  exist after fixing the K\"ahler moduli.
  They say that  suitable corrections to the superpotential
depending on the K\"ahler moduli will not materialize.
This means that string consistency conditions, like absence of Freed-Witten
 anomalies, correct number of instanton 0-modes etc., could
 generically forbid any such instanton correction.
  Here we proceed under the
  usual assumption that such an instanton exists and derive its
  consequences. This will eventually  lead us to the conclusion that the 
  uplifted de Sitter minimum of the DKMM-refined KKLT scenario is in
  the swampland. Thus, logically  our result is consistent with \cite{Lust:2022lfc}.

\subsection{Mass scales}

The position $\tau$ of the supersymmetric AdS minimum is determined by the solution of the equation
\eq{
    \label{taumin}
    \abs{A} (2a\tau+3)=3|W_0| e^{a\tau}
}  
yielding the value of the cosmological constant
\eq{
  V_{\rm AdS}\sim -{g_s\over \tau^3} |W_0|^2\, M_{\rm pl}^4\,.
}  
For a successful uplift to a dS vacuum, this has to be of the same
scale as the $\ov{D3}$-brane tension
\eq{
  \label{Vuplift}
  V_{\rm up}\sim {1\over g_s M^2} \left({|Z|\over {\cal
        V}_w}\right)^{\!{4\over 3}} \, M_{\rm pl}^4\,.
}
This leads to
\eq{
 \label{upliftcond} 
  |W_0|\sim  {1\over g_s \vert M\vert} \tau^{1\over 2} |Z|^{2\over 3}\,.
 } 
We note that  for an exponentially small value of $Z$,
this seems to be parametrically compatible with the race-track
condition ${|W_0| > g_{s} M^2 |Z|/\left( \left(2 \pi\right)^2|K|\right)}$.
Indeed, combining the latter with \eqref{upliftcond} leads to
\eq{
  \label{Zinequal}
    \left( {|Z|\over {\cal V}_w} \right)^{1 \over 3} \lesssim {\left(2\pi \right)^2N \over (g_s M^2)^2} \,,
}
which, however, is parametrically saturated in Scenario 1.
In this case,
the exponentially small $|Z|$ has to be parametrically equal to
a polynomially small parameter. This  might appear fairly unnatural
and certainly requires a large value of the control parameter $g_s M^2$. We will see that it
implies severe problems with the validity of the employed low energy effective action in
the warped throat.

Moreover,  we notice that the bound \eqref{Zinequal} is  compatible with control over the
warped effective action, i.e. with the  upper bound
\eqref{controlcond} if \footnote{The condition \eqref{controlcond2} is a necessary (sufficient) condition in Scenario 1 (2) for a control over the warped effective action. In the following we will impose the condition \eqref{controlcond2} in both Scenario 1 and Scenario 2 to guarantee the control over the effective theory.}
\eq{
    \label{controlcond2}
     N\lesssim  {1\over (2\pi)^2} \left(g_s M^2\right)^{3\over 2}\,.
}
Let us proceed with the determination of the remaining parameters. For an exponentially small value of $Z$, one can estimate the value of $\tau$ as
\eq{
  \label{fixatau}
       a\tau\sim -\log |W_0|
     \lesssim -\log |Z|\sim {2 \pi N\over g_s M^2}\,.
}
Invoking the condition \eqref{werder}, namely $\tau> 2\pi N$, one gets in
both Scenario 1 and Scenario 2
\eq{
  \label{acondition}
 \boxed{ a\lesssim  {1\over g_s M^2} \ll 1 }
}  
which, as already observed in \cite{Carta:2019rhx}, is violated for an isotropic CY
and a D3-brane instanton, i.e. $N_c=1$. The condition \eqref{acondition}
must also hold in order to avoid the singular-bulk problem discussed
in \cite{Gao:2020xqh}.
We notice that the string frame volume of the 4-cycle is given by $\hat\tau\gtrsim
g_s N = (g_s \vert M\vert) \vert K\vert>1$ and therefore guaranteed to be  in the large volume regime.

The mass of the K\"ahler modulus scales as
\eq{
          \label{masskaehler}
          m_T\sim a{ g_s^{1\over 2}\over {\cal V}_w^{1\over 3}} |W_0|\, M_{\rm pl} 
}
which, as already mentioned,  scales in the same way with $|W_0|$ as the mass of the light $S$
modulus. Therefore, one might suspect that first integrating out $S$ is not
 self-consistent. It was argued in \cite{Demirtas:2021nlu,Demirtas:2021ote} that as long as the
parameter $A$ in the KKLT superpotential does not depend on any
complex structure modulus, the minimum prevails. We checked this
numerically in concrete examples.

\subsection{Mass hierarchies extended}
\label{Mass hierarchies extended}

Now we would like to see how the mass scale of the uplift potential fits
into the hierarchies of scales in \eqref{masshierarchya}.
Using  \eqref{Vuplift} we can write
\eq{
     V_{\rm up}^{1\over 4}\sim {\Lambda\over (g_s M^2)^{3\over 4}}\,                    
}
which  is  smaller than the species scale and larger than the mass of
$Z$.

Next we analyze the relation between the mass of the K\"ahler modulus
and the light KK scale. Using the relations
\eqref{upliftcond} and \eqref{fixatau} one gets 
\eq{
  \label{masstau}
  {m_T } \lesssim {2 \pi N\over (g_s M^2)^{3\over 2}}  \left( {|Z|\over {\cal
        V}_w}\right)^{\!{2\over 3}} M_{\rm pl}\sim {2 \pi N\over (g_s M^2)^{2}}  \left( {|Z|\over {\cal
        V}_w}\right)^{\!{1\over 3}}\, \Lambda\,.
}
Thus, one finds
\eq{
  \label{ratiomasstaukk}
  {m_T\over m_{\rm KK}}\lesssim {2 \pi N\over (g_s M^2)} \left( {|Z|\over {\cal
  V}_w}\right)^{\!{1\over 3}} \lesssim {\left(2 \pi\right)^3 N^2 \over (g_s M^2)^3} \lesssim 1 \,,
}
where we have used the relation \eqref{Zinequal} in the second step and the constraint \eqref{controlcond2} needed for the control of the effective theory in the last step. Hence, we have the desired hierarchy
$m_T<m_{\rm KK}$.

Let us perform a similar analysis for  the relation between the mass of the $S$ modulus and  the light KK scale. Using the relation \eqref{upliftcond} one gets for Scenario 2
\eq{
    {m^{(2)}_S \over m_{\rm KK}}\sim {c_1 c_2 \over g_s^2 } \left({|Z|\over {\cal V}_w}\right)^{\!{1\over 3}} \lesssim {\left(2 \pi\right)^2 c_1 c_2  N \over(g_s M)^4} \,,
}
where we have used the relation \eqref{Zinequal}  in the last step. Hence a sufficient condition for having the desired hierarchy $m^{(2)}_S<m_{\rm KK}\sim m_Z$ is
\eq{
  \label{Nbound_2}
  \boxed{\left(2 \pi\right)^2 c_1 c_2 N < (g_s M)^4}\,.
}
As already mentioned in section \ref{Conifold modulus and axio-dilaton}, we find in Scenario 1 the ratio
\begin{equation}
{ m^{(1)}_S  \over m_Z } = {N \over \tau} |Z|^{2\over 3} \ll 1\,,
\end{equation}
where we have used that $\tau > N$. Hence, we have the desired hierarchy $m^{(1)}_S<m_{\rm KK}\sim m_Z$ in Scenario 1.
Note that  in Scenario 1, for the ratio of the two lightest moduli we find 
\eq{
{m_T\over m^{(1)}_S }\sim {g_s M^2\over 2\pi N} \lesssim 1\,,
}
where the inequality in the second step has to hold to guarantee $|Z| \ll 1$.

Summarizing, our analytic analysis revealed  the following order of
all the relevant red-shifted mass scales of the warped throat 
\begin{equation}
\label{chain_ineq}
    \Lambda > \tilde{\Lambda}_{\text{sp}} > V^{1\over 4}_{\text{up}} >m_{Z}\sim m_{\rm
      KK}  > m_{S/T}\,,
\end{equation}
where we have the hierarchy $m^{(1)}_S > m_T$ in Scenario 1 and there is no definite hierarchy between the masses $m^{(2)}_S$ and $m_T$ in Scenario 2.

\subsubsection*{Mass scales in Scenario 1}

Let us first consider the mass scales in Scenario 1 in more detail.
Upon invoking the
parametrically saturated relation \eqref{Zinequal},
the cut-off can be expressed as
\eq{\Lambda \sim { \left(2 \pi\right)^2 N\over (g_s
    M^2)^{3\over 2}}   M_{\text{pl}}\,.
}
We have seen that the masses in the throat are
ordered in the expected manner. However, using the saturated relation
\eqref{Zinequal}, one can now express the lightest bulk mass scale
\eqref{masscomplexstr} as
\eq{
    \label{M_cs}
    M_{\rm cs} \sim g_s^{1 \over 2} { (g_s M^2)^{3 \over 2} \over \left(2 \pi\right)^2 N \tau^{3\over 2}} \Lambda \lesssim {g_s^{1 \over 2} \over 2\pi} {\left( g_{s} M^2  \right)^{3 \over 2}\over \left( 2 \pi N \right)^{5\over 2} } \Lambda\,
}
which implies
\eq{
    M_{\rm cs} \lesssim   {g_s^{1 \over 2} \over 2\pi} \left({g_s M^2\over
    2\pi N}\right)^{5 \over 2} m_{\rm KK} \lesssim m_{\rm KK}\,.
 }
This means that the bulk complex structure moduli are lighter 
than the  red-shifted KK-modes in the throat. Hence, in Scenario 1 the bulk and
throat are not energetically separated, spoiling completely
the validity of the used effective action in the throat.
One might think that this is to be expected from the very begining as we
were balancing a term in the IR (throat)  against a term in the UV (bulk).
However, notice that the tree-level UV term is vanishing by
construction and that in 
the racetrack potential  \eqref{racetrack} we were  balancing the IR energy against a
non-perturbative bulk term. That this spoils  decoupling of bulk and
throat is not obvious.

Next we analyze the relation between the lightest bulk mass scale $M_{\rm cs}$ and the mass of the $S$ modulus. Using the saturated relation \eqref{Zinequal}, we get the ratio
\eq{
    { M_{\rm cs} \over m_S^{(1)}} \sim {g_s^{1 \over 2}\over \tau^{3\over 2} }\, {\left( g_s M^2 \right)^{13\over 2} \over \left(2 \pi\right)^6 N^4}   \,
}
so that imposing  $M_{\rm cs}>m_S^{(1)}$ sets an
upper bound on $\tau$.

Let us roughly estimate the number of bulk complex structure moduli
to be ${h_{31}\sim \chi(Y)\gtrsim N}$, where in the final step we have assumed
that the large tadpole from the fluxes in the throat $N=MK$ is not cancelled
by other contributions.
Since these extra fields and all  their KK-modes are now light,
the species scale also becomes smaller and can
be estimated as
\eq{  \label{species scale scenario1}
                 \tilde\Lambda_{\text{sp}} \lesssim {\Lambda\over N^{1\over 2} (g_s
                   M^2)^{1\over 3}}\,.
}
Apparently, this scale is smaller than $V_{\rm up}^{1\over 4}$ as
\eq{
                   \tilde\Lambda_{\text{sp}} \lesssim \left({g_s M^2\over
                 N}\right)^{1\over 2} V_{\rm up}^{1\over 4} \lesssim V_{\rm up}^{1\over 4}\,.
}
We notice that for this more specific model we have arrived at a very
similar conclusion as \cite{Lust:2022lfc},
though not using holography but the existence of an uplift of the
initial AdS-minimum. 
In table \ref{tab_allmass} we show the hierarchy of the bulk and the
throat mass scales. On the left we list the expected order, if
there were a separation between the bulk and the red-shifted throat.
On the right, we list the non-separated mass scales found
for Scenario 1.

\begin{table}[ht] 
  \renewcommand{\arraystretch}{1.5} 
  \begin{center} 
    \begin{tabular}{|c|} 
      \hline
      bulk-throat separated scales       \\
      \hline \hline
      $M_{\rm cs}\sim  {\sqrt{g_s} \over  {\cal V}_w} M_{\rm pl}$ \\[0.075cm]
      \hline
      $\Lambda \sim M_{\text{pl}} \eta^{3\over 2} N$ \\[0.075cm]
      \hline
      $\tilde\Lambda_{\text{sp}}\sim  \eta^\frac{1}{3} \Lambda$\\[0.075cm]
      \hline     
      $V_{\rm up}^{1\over 4}\sim \eta^\frac{3}{4}  \Lambda$\\[0.075cm]
      \hline
      $m_Z\sim m_{\rm KK}\sim \eta \Lambda$ \\[0.075cm]
      \hline
      $m_S^{(1)}\sim   N^3 \, \eta^{5}    \Lambda $ \\[0.075cm]
      \hline  
      $m_T\sim  N^2\, \eta^{4}    \Lambda $ \\[0.075cm]
      \hline
      \end{tabular}
      \hspace{0.8cm}
      \begin{tabular}{|c|} 
      \hline
      bulk-throat mixed scales     \\
      \hline \hline
      $\Lambda \sim M_{\text{pl}} \eta^{3\over 2} N$ \\[0.075cm]
      \hline  
      $V_{\rm up}^{1\over 4}\sim \eta^\frac{3}{4}  \Lambda$\\[0.075cm]
      \hline
      $\tilde\Lambda_{\text{sp}} \lesssim  N^{-{1\over 2}}  \eta^\frac{1}{3} \Lambda$\\[0.075cm]
      \hline   
      $m_Z\sim m_{\rm KK}\sim \eta \Lambda$ \\[0.075cm]
      \hline
      $M_{\rm cs} \lesssim  g_s^{1\over2} N^{-{5\over 2}}  \eta^{-{3 \over 2}} \Lambda$\\[0.075cm]
      \hline
      $m_S^{(1)}\sim  N^3\, \eta^{5}    \Lambda$ \\[0.075cm]
      \hline  
      $m_T\sim  N^2 \, \eta^{4}    \Lambda $
      \\[0.075cm]
      \hline
      \end{tabular}
      \caption{Expected separated and realistic mixed mass scales for Scenario 1. We have defined the parameter  $\eta =1/ \left(g_s M^2\right)$.}
    \label{tab_allmass} 
  \end{center} 
\end{table}

In addition, we have  a very large  tadpole contribution from the fluxes. From the now saturated relation \eqref{Zinequal}, we can derive the lower bound
\eq{
  \label{boundonZ}
  N \gtrsim 2\pi (\log |Z|)^4 |Z|^{-{2\over3}}\,,
}
where we also used \eqref{werder}. Remarkably, the bound is solely determined by the VEV of the conifold modulus.
To get a better impression, let us assume that for having control over the geometry and the effective action we need the conifold modulus in a regime $|Z|<10^{-5}$. Then equation
\eqref{boundonZ} tells us that $N > 10^8$, which implies $M\sim
10^{4}$ and $a<10^{-7}$ (for $M\approx N^{1\over 2}$ and $g_s \sim
0.1$). For $|Z|<10^{-4}$ one finds $N > 10^7$.

Thus, in addition to having the wrong hierarchy we face large integers, i.e. large fluxes and numbers of branes, large tadpole contributions and either large non-isotropies in the bulk or a high rank confining gauge group.

\subsubsection*{Mass scales in Scenario 2}

This case is less constrained, as we treat the parameters $c_1$ and
$c_2$ in the race-track potential as free parameters. Thus the
constraint \eqref{Zinequal} is mild now. Indeed, the lightest bulk mass scale $M_{\rm cs}$ is larger than the cut-off $\Lambda$, if
\eq{ 
     \left(  {\cal
        V}_w^2 \vert Z\vert \right)^{1/3}\lesssim \frac{1}{\vert M\vert}\,.
}
In this case the bulk and the throat are energetically decoupled and we have the following hierarchies of scales
\begin{equation}
\label{chain_ineq_Scenario2}
    M_{\rm cs} > \Lambda > \tilde{\Lambda}_{\text{sp}} > V^{1\over 4}_{\text{up}} >m_{Z}\sim m_{\rm
      KK}  > m_{S/T}\,.
\end{equation}
As a proof of principle, in the next section we provide an example
that features the intended hierarchy \eqref{chain_ineq_Scenario2} of all masses.

However, we can still derive a lower bound on the
  fluxes in the throat.
  From the uplift condition \eqref{upliftcond} and  the relations \eqref{approximate_W0}
and \eqref{Z_minimum} it follows that
\eq{ \label{intermediate_rel}
    {\vert c_{2}-c_{1}\vert \over c_{2}}\, \vert a_{1}\vert \,e^{-c_{1} s_{0}}\sim 
    {1\over g_s \vert M\vert} \tau^{1\over 2} |\zeta_0|^{ 2\over 3} e^{-{2\over 3}{2\pi K\over  M}s_0}\,.
}
Setting the exponentially small terms in \eqref{intermediate_rel} to be of the same scale, we can estimate
\eq{
  \label{estimate_c1}
  c_1\sim {4\pi K\over 3 M}\,.
}
Invoking now the (slightly concretized) bound $g_s |M|\gtrsim O(10^n)$
from \eqref{sugraregime}, set by
having a valid supergravity, large radius description, we get 
\eq{
    M \gtrsim {10^n\over g_s}\,,\qquad
    K \gtrsim {10^n c_1 \over g_s}\,,\qquad 
    N \gtrsim {10^{2n} c_1 \over g_s^2}\,.
}
Assuming $c_1 > g_s$  for a controlled race-track minimum,
$g_s\lesssim 0.1$ for control over string loop corrections and 
$g_s |M|\gtrsim  10$ for a controlled supergravity description, one finds $N\gtrsim 10^{2-3}$. This is still a moderately large contribution that might be in conflict with tadpole cancellation or the tadpole conjecture, respectively. For fully fledged models we expect the bound to be much more stringent, as we treated the race-track parameters as free parameters while for concrete Calabi-Yau manifolds they are also determined by three-form fluxes contributing
to the tadpole themselves. Hence, the tadpole will easily exceed the above simple estimate by orders of magnitude.

\subsubsection*{Comments on tadpole cancellation}

Recall that  the tadpole cancellation condition reads
\eq{
    MK + N_{\rm flux}+N_{\rm brane}={\chi(Y)\over 24}
}
where $N_{\rm flux}$ denotes the contribution of the other present
fluxes and $N_{\rm brane}$ the contribution from D3-branes and
magnetized D7-branes. For such a large  flux $N=M K$ one does not
expect (almost fine tuned) cancellations between the various flux and brane contributions to occur. Therefore, one needs a Calabi-Yau fourfold with a very large Euler characteristic $\chi(Y)$. This will have many complex structure moduli so that one very likely encounters the problem of the tadpole conjecture
\cite{Bena:2020xrh,Bena:2021wyr,Plauschinn:2021hkp,Grana:2022dfw}.
Namely, that it is not possible to freeze
all of these many moduli using three-form fluxes, something that was
silently assumed before we focused just on the final two moduli $Z$ and $S$. Hence, we arrive at the conclusion that 
a working DKMM-refined KKLT Scenario 1 would require very large fluxes that are in conflict with the tadpole constraint. Scenario 2 is expected to be much more constrained in concrete cases, so that the flux tadpole could become dangerously large, likewise. Note that a very similar conclusion was drawn recently for a controlled Large Volume Scenario \cite{Junghans:2022exo,Gao:2022uop,Junghans:2022kxg}.


\section{Numerical analysis}
\label{Numerical analysis}

Since throughout our analysis we were invoking quite a number of approximations, we need to provide a proof of principle. For this purpose, let us now test the previous results by comparing them with concrete numerical solutions of the problem.

We consider the full scalar potential after analytically stabilizing the conifold modulus and systematically search for a region in parameter space with a dS minimum by making use of the uplift condition, which will be written as
\eq{
    \label{uplifteq}
    \kappa {g_s |W_0|^2\over \tau_0^3} \sim  {\xi\over g_s
    M^2}  {|Z|^{4\over 3}\over \tau_0^2} 
}
where $\xi$ is the numerical prefactor\footnote{From \eqref{Vantitot} we could read off $\xi={9\over 8}c''\simeq  {63\over 32}$ but here we prefer to treat it as a parameter.} appearing in the uplift potential and we introduced an order one ``balance'' parameter $\kappa$. Since order one factors do matter in this analysis, we have at least taken factors of $(2\pi)$ into account. In hindsight, utilizing  results from moduli stabilization we derive a condition on the fluxes $(M,K,M_1)$ for a successful uplift.

The construction of a working uplift turns out to be quite challenging if the free race-track parameters are all dialed by hand. Thus, we make specific assumptions for some coefficients to simplify our search. Recall that after the stabilization of the conifold modulus, the generic superpotential reads
\eq{
    W = a_{1}e^{-c_{1}S} + a_{2}e^{-c_{2}S}
}
and its value at the minimum can be approximated by
\eq{
    W_{0} \sim {c_{2}-c_{1} \over c_{2}} a_{1} e^{-c_{1} S_{0}}
}
with $S_{0}$ given by \eqref{smin} as usual. Motivated
by the race-track scenario with non-perturbative terms generated via gaugino condensation, we now assume $c_{i} <1$ and that the second exponent in the superpotential is given in terms of the first one via\footnote{Note that for gaugino condensation, $c_1=1/(N+1)$ and $c_2=1/N$ are
satisfying \eqref{racegaug}.}
\eq{
  \label{racegaug}
    c_{2} = (c_1^{-1}-1)^{-1}\,.
}
The factor $a_{2}$ can be adjusted such that the log-term in \eqref{smin} is a number of order one. With these premises we derive the following expressions for $W_{0}$ and the string coupling $g_{s}$:
\eq{
    W_{0} \sim a_1 c_1 \exp \left( -\frac{1}{c_1} \right) \,, \qquad g_{s} \sim c_1^2\,.
  }
Thus, this ansatz guarantees that we have control over the race-track potential as $g_s\sim c_1^2 < c_1 $ for $c_1<1$.  

Moreover, $W_0$ allows us to estimate the value of $\tau$ via \eqref{fixatau}. To improve the quality of our estimate the log-log correction will be taken into account, meaning that
\eq{
    \label{estimate_Tau}
    a \tau \sim \log{\abs{W_{0}}^{-1}} + \log{\log{\abs{W_{0}}^{-1}}}.
}
This effectively increases our estimate for $a\tau$ in both scenarios. Respecting the condition $\tau>2 \pi N$, we can express it as
\eq{
    \label{newacondition}
    a = {\lambda \over 2\pi N}\left(\log{\abs{W_{0}}^{-1}} + \log{\log{\abs{W_{0}}^{-1}}}\right)
}
where we introduced a factor $ 0<\lambda<1$, which for each concrete numerical example allows some tuning. The Pfaffian $A$ is taken to be $-1$ and the coefficient $\xi$ is taken to be $\frac{63}{32}$  in all concrete realizations presented below. For later purposes, we define the two parameters
\eq{
    \label{XYdef}
    X = {M \over 2\pi K}\,, \qquad\quad Y = 2\pi \vert K\vert\,. 
}


\subsection{Scenario 1}
\label{num1}

In Scenario 1 we have\footnote{We got rid of the ${1 / i}$ factor in $a_1$ (c.f. $W_{\rm eff}$ in \eqref{racetrack}) by shifting the axion $C_0$.} $a_{1} = M\zeta_0 / (2\pi)$ and $c_{1} =
2\pi K/M $, so the conifold modulus and the superpotential are given by
\eq{
  Z\sim \zeta_0\exp\left(-{M \over 2\pi K} \right)\,,\qquad
  W_{0} \sim K \zeta_0 \exp\left(-{M \over 2\pi K}\right)
}
and the uplift condition \eqref{uplifteq} can be expressed as
\eq{
    \label{uplift constraint scneario1}
    {\xi \over \kappa} {2 \pi  \over Y^2} X^3 \left\vert \zeta_0\right\vert^{-2/3}\exp\left( {2X \over 3}\right) \simeq \lambda\,.
}
For really finding parameters allowing a controllable dS uplift, this condition needs to be satisfied with a sufficient accuracy. Let us stress that \eqref{newacondition} leads to  the following refinement of the condition \eqref{acondition} 
\eq{
    a\lesssim {2\pi\over g_s M^2} + \frac{1}{N}\log\left({2 \pi \over g_s \vert M\vert \left\vert \zeta_0\right\vert}  \right) \ll 1 \,.
}
As a consequence, the search for a numerical realization becomes a lot easier.

Employing the above recipe, we were able to construct numerical realizations
with a tadpole of $O(10^{19})$ induced by the throat fluxes, which
surpasses the expected lower bound \eqref{boundonZ} by many orders of
magnitude. These examples also confirm the unphysical mass hierarchy
with mixed throat and bulk scales. Our large flux numbers are
especially enforced by the stabilized $\tau$-modulus. Indeed, the smallness of
the $a$ coefficient and the uplift condition impose such a large value
for $\tau$, as  the strong warping condition \eqref{strongwarping}
turns out to be  violated below $N \sim O(10^{19})$.
It may be possible to get closer
to the theoretical lower bound of $N \sim 10^{7-8}$ in more finely
tuned setups, but in any case the flux tadpole of Scenario 1 has
certainly almost no chance to be cancelled in realistic situations.

\subsection{Scenario 2}
\label{num2}

In the second scenario, the race-track coefficients are not determined by the  flux numbers $M,K,M_1$ and
we dial them freely as long as we stay in the controlled regime. However, keep in  mind that in concrete realizations those coefficients are not free parameters but also set by other flux numbers and data of  the underlying Calabi-Yau manifold. Therefore, the way we proceed is very optimistic and eventually Scenario 2 will also be much more constrained. Setting $a_{1} = 1$ for convenience, we can derive the following uplift constraint for Scenario 2:
\eq{
    \label{uplift constraint scneario2}
    {\xi \over \kappa}  (2\pi X)^{-1}  c_{1}^{-6} \vert\zeta_0 \vert^{\frac{4}{3} } \exp\left({2 \over c_{1}} -\frac{4}{3}  \frac{1}{c_{1}^2 X} \right) \simeq \lambda\,.
    }

One numerical example is specified by $c_{1} = 1/40$, $\lambda = 9/10$, $\kappa = 1/2$, $M = 5\,000$, ${K = 37}$ and $M_1=\frac{M}{2\pi i}$. The coefficient $a$ is therefore $a\approx 3.67 \cdot 10^{-5} $. The D3-brane tadpole contribution is of order ${N=O(10^{5})}$, which is smaller
than in Scenario 1 but still fairly large. For $s\rvert_{\text{num}} \approx 1\,599.75$ and $\tau\rvert_{\text{num}} \approx 1.30 \cdot 10^6$ the dS minimum is found at
\eq{
    V_{0} \rvert_{\text{num}} \approx 5.88 \cdot 10^{-60} M_{\text{pl}}^4 \,.
}
In table \ref{scales_scenario2} we show the numerical values of the mass scales in the left-hand column. The numerical values of the mass scale of the bulk complex structure moduli and the scales $\Lambda$ and $\tilde\Lambda_{\text{sp}}$ were determined via the relations \eqref{masscomplexstr}, \eqref{cut-off_Lambda} and \eqref{species_scale} respectively. $V_{\rm up}^{1\over 4}$ and the mass scale $m_{\rm KK}$  were obtained via \eqref{Vantitot} and  \eqref{m_KK} respectively. The masses of the lightest saxions $s$ and $\tau$ were obtained via \eqref{eq:massmatrix}.

\begin{table}[ht] 
  \renewcommand{\arraystretch}{1.5} 
  \begin{center}
    \begin{tabular}{|c|c|} 
      \hline
      numerical values   &  theoretical predictions    \\
      \hline \hline
       $M_{\rm cs}\approx 1.69 \cdot 10^{-11} M_{\text{pl}}$ & $M_{\rm cs}\approx 1.71 \cdot 10^{-11} M_{\text{pl}}$ \\[0.1cm]
      \hline 
       $\Lambda\approx 1.87 \cdot 10^{-12}  {M_{\rm pl}}$ & $\Lambda\approx 1.87 \cdot 10^{-12}  {M_{\rm pl}}$ \\[0.1cm]
     \hline
     $\tilde\Lambda_{\text{sp}} \approx  7.50 \cdot 10^{-14}M_{\text{pl}}$ & $\tilde\Lambda_{\text{sp}} \approx    7.50 \cdot 10^{-14}M_{\text{pl}}$\\[0.1cm]
  \hline   
  $V_{\rm up}^{1\over 4}\approx 1.59\cdot 10^{-15}M_{\text{pl}}$ & $V_{\rm up}^{1\over 4}\approx 1.59 \cdot 10^{-15}M_{\text{pl}}$\\[0.1cm]
     \hline
      $m_Z\sim m_{\rm KK}\approx 1.20 \cdot 10^{-16} M_{\text{pl}}$ & $m_Z\sim m_{\rm KK}\approx 1.20 \cdot 10^{-16} M_{\text{pl}}$ \\[0.1cm]
      \hline
     $ m_s \approx 3.00 \cdot 10^{-27}M_{\text{pl}}$ & $m_s\approx 2.99 \cdot 10^{-27}M_{\text{pl}}$\\[0.1cm]
      \hline
     $ m_\tau\approx 2.10 \cdot 10^{-29} M_{\text{pl}} $ & $ m_\tau \approx 8.59 \cdot 10^{-29}M_{\text{pl}} $\\[0.1cm]
      \hline 
    \end{tabular}
    \caption{The numerical values and the theoretical predictions of the mass scales for Scenario 2.      }
    \label{scales_scenario2} 
  \end{center} 
\end{table}
  
This has to be compared with our  theoretical predictions. Here we find the dS minimum at  $s\rvert_{\text{theo}} \approx 1\,600.00$ and $\tau\rvert_{\text{theo}} \approx 1.29 \cdot 10^6$ with the value of the potential being
\eq{
    V_{0} \rvert_{\text{theo}} \approx 5.98 \cdot 10^{-60} M_{\text{pl}}^4\,.
}
In the right-hand column of table \ref{scales_scenario2}, we list the
theoretical predictions of the mass scales, using the same formulas as
those used for the numerical values, except for the masses of the
lightest saxions, whose theoretical values were obtained with the
approximate formulas \eqref{massdilaton} and \eqref{masskaehler}.

\begin{figure}[!ht]
\centering
\includegraphics[width =\textwidth] {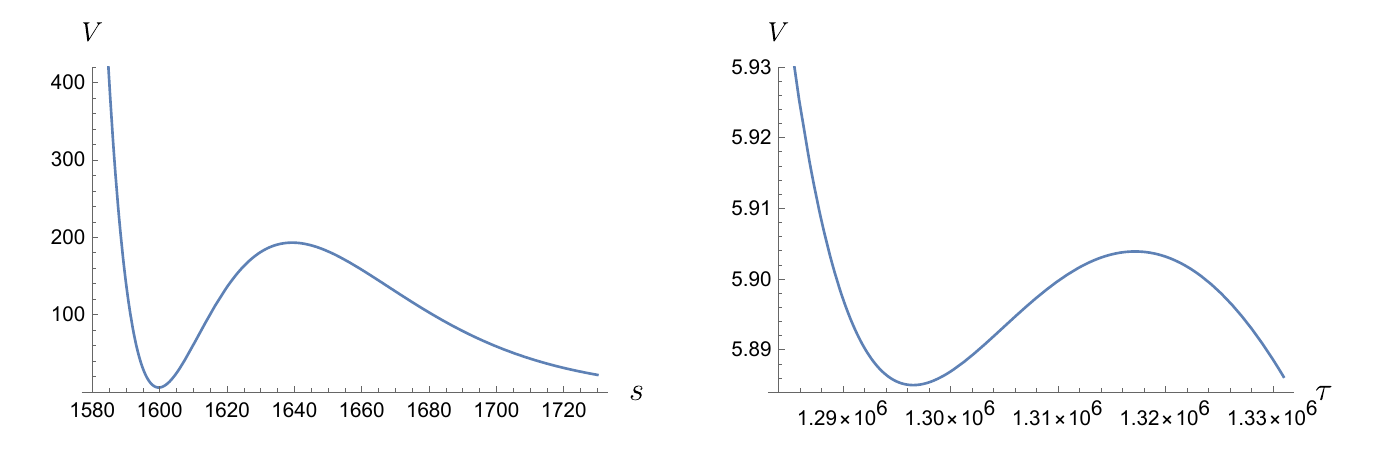}
\caption{Plots of scalar potential showing a realization of Scenario 2. Left: $V(s,\tau)$ for $\tau \approx 1.30 \cdot 10^6$ and right: $V(s,\tau)$ for $s \approx 1\,599.75$, both multiplied by $10^{60}$. Choice of parameters: $c_{1} = 1/40$, $\lambda = 9/10$, $\kappa = 1/2$, $M = 5\,000$, $K = 37$, $M_1=\frac{M}{2\pi i}$.} 
\label{kklt_scenario2_2d}
\end{figure}

\begin{figure}[!ht]
\centering
\includegraphics[width =.75 \textwidth]{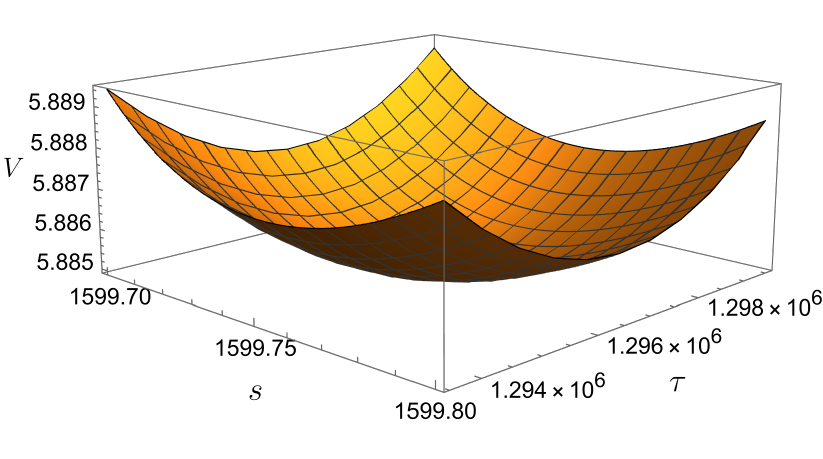}
\caption{Potential $V(s,\tau)$ for our example from Scenario 2, multiplied by $10^{60}$. Choice of parameters: $c_{1} = 1/40$, $\lambda = 9/10$, $\kappa = 1/2$, $M = 5\,000$, $K = 37$, $M_1=\frac{M}{2\pi i}$.} 
\label{kklt_scenario2_3d}
\end{figure}

Hence, the theoretical predictions for the mass scales are in agreement with the numerical results. Again we show plots of the potential close to this minimum in the
figures \ref{kklt_scenario2_2d} and \ref{kklt_scenario2_3d}.


This example for Scenario 2 indicates that while the hierarchy of mass scales is in order, the fluxes $M,K$ needed to stabilize the conifold modulus and the axio-dilaton yield a moderately large contribution
$N=M K$ to the D3-brane tadpole. By choosing different values for the parameters one might come even closer to the general bound $N>10^{3-4}$.


\section{Conclusion}
\label{sec_con}

In this paper we extended the usual KKLT construction
by the concrete mechanism of DKMM to stabilize
the complex structure moduli such that $W_0$ is guaranteed to be exponentially
small. The objective was to derive what additional constraints
this imposes on the length and mass scales of the geometry
and the light fields involved.

First, using  geometric consistency
constraints we derived a lower and an upper bound for the 
length of the throat. Mutual consistency then directly led us to a bound for the
four-cycle volume  in terms of the D3-tadpole contribution
coming from the throat fluxes. Thus, this direct computation for the
Klebanov-Strassler throat  is consistent with a former indirect
argument using the D3-brane backreaction \cite{Carta:2019rhx}.

We then considered the explicit stabilization of the final light moduli
$(Z,S,T)$, working in a framework where all heavier complex structure
moduli have been assumed to be fixed via three-form fluxes. We employed the
methods and the analysis of \cite{Blumenhagen:2019qcg}, but generalized it such
that we also  included  the axio-dilaton and therefore the string
coupling constant in the set of light fields. Following the
generalization
of the DKMM mechanism to the conifold regime 
we invoked a race-track scenario for its stabilization.
For the absence of non-perturbative corrections in the K\"ahler moduli, the scalar potential was found to be
positive-semidefinite and to obey a modified no-scale structure.

Computing the appearing mass scales, we found that this
  DKMM-refined  KKLT scenario comes with a couple of strong constraints.
In our more restricted Scenario 1 we found that one is driven to a
regime where the bulk mass scale does not decouple from the throat mass
scale. Clearly, this invalidates the employed low energy effective
action in the throat. Moreover, analogous to the recent work
\cite{Lust:2022lfc} we found that as a consequence, the
energy scale of the uplift
is parametrically larger than the
(throat) species scale.
Recall that \cite{Lust:2022lfc}
claim  that the whole DKMM construction ceases to
be controllable even before the uplift. 

We also found that for both Scenario 1 and Scenario 2,
the two flux quanta $M,K$
need to be fairly large, in fact much larger than what was derived
from the non-destabilization of the conifold modulus in
\cite{Bena:2018fqc}. Moreover, also the parameter $a$  that appears
in the combination $aT$ in the non-perturbative term needs
to be extremely small, which requires either a high rank confining
gauge group or a highly non-isotropic 4-cycle. All this was confirmed
by  concrete numerical examples, whose
throat mass scales were consistent with our theoretical predictions.
In view of the recent tadpole conjecture,  this raises some
additional doubts that   the DKMM-refined KKLT scenario
is in the landscape of string theory.

\vspace{0.3cm}
\noindent
\subsubsection*{Acknowledgments}
We would like to thank Andriana Makridou and Erik Plauschinn for
helpful discussions. We are also grateful to Arthur Hebecker, Daniel
Junghans and Severin L\"ust for clarifying comments about a former version of this paper.

\vspace{1cm}

\clearpage

\bibliography{references}  
\bibliographystyle{utphys}


\end{document}